\documentclass[a4paper,twocolumn,11pt,accepted=2022-07-30]{quantumarticle}
\pdfoutput=1
\usepackage[utf8]{inputenc}
\usepackage[english]{babel}
\usepackage[T1]{fontenc}
\usepackage{hyperref}
\usepackage{bm,bbold}
\usepackage{graphicx}
\usepackage{bbm}
\usepackage{color}
\usepackage{enumerate}
\usepackage{amsfonts, amsmath, amsthm, amssymb} 
\usepackage{slashbox}
\usepackage{comment}
\usepackage[normalem]{ulem}

\def \diracspacing {0.7pt}
\newcommand{\bra}[1]{\langle #1 \hspace{\diracspacing} |} % bra
\newcommand{\ket}[1]{| \hspace{\diracspacing} #1 \rangle} % ket
\newcommand{\braket}[2]{\langle #1 \hspace{\diracspacing} | \hspace{\diracspacing} #2 \rangle} % braket with different vectors
\newcommand{\braketq}[1]{\braket{#1}{#1}} % braket with the same vector
\newcommand{\ketbra}[2]{| \hspace{\diracspacing} #1 \rangle \langle #2 \hspace{\diracspacing} |} % ketbra with different vectors
\newcommand{\ketbraq}[1]{\ketbra{#1}{#1}} % ketbra with the same vector
\newcommand{\I}{\mathbb{1}}
\newcommand{\ii}{\mathrm{i}}
\newcommand{\ee}{\mathrm{e}}
\newcommand{\bx}{\mathbf{x}}

\DeclareMathOperator{\tr}{tr}
\DeclareMathOperator{\diag}{diag}

\theoremstyle{definition}
\newtheorem{defn}{Definition}%[section]
\theoremstyle{plain}

\newtheorem{prop}[defn]{Proposition}

\begin{document}

\title{Three numerical approaches to find mutually unbiased bases using Bell inequalities}

\author{Maria Prat Colomer}
\altaffiliation{These authors contributed equally to this work.}
\affiliation{ICFO-Institut de Ciencies Fotoniques, The Barcelona Institute of Science and Technology, 08860 Castelldefels, Spain}
\affiliation{CFIS-Centre de Formació Interdisciplinària Superior, UPC-Universitat Politècnica de Catalunya, 08028 Barcelona, Spain}
\orcid{0000-0002-7866-8356}

\author{Luke Mortimer}
\altaffiliation{These authors contributed equally to this work.}
\affiliation{ICFO-Institut de Ciencies Fotoniques, The Barcelona Institute of Science and Technology, 08860 Castelldefels, Spain}
\orcid{0000-0002-5644-8985}

\author{Irénée Frérot}
\affiliation{ICFO-Institut de Ciencies Fotoniques, The Barcelona Institute of Science and Technology, 08860 Castelldefels, Spain}
\affiliation{Univ Grenoble Alpes, CNRS, Grenoble INP, Institut N{\'e}el, 38000 Grenoble, France}
\orcid{0000-0002-7703-8539}

\author{Máté Farkas}
\email{mate.farkas@icfo.eu}
\affiliation{ICFO-Institut de Ciencies Fotoniques, The Barcelona Institute of Science and Technology, 08860 Castelldefels, Spain}
\orcid{0000-0002-2682-8215}

\author{Antonio Acín}
\affiliation{ICFO-Institut de Ciencies Fotoniques, The Barcelona Institute of Science and Technology, 08860 Castelldefels, Spain}
\affiliation{ICREA-Institucio Catalana de Recerca i Estudis Avançats, Lluis Companys 23, 08010 Barcelona, Spain}
\orcid{0000-0002-1355-3435}

%\date{\today}

\begin{abstract}
Mutually unbiased bases correspond to highly useful pairs of measurements in quantum information theory. In the smallest composite dimension, six, it is known that between three and seven mutually unbiased bases exist, with a decades-old conjecture, known as Zauner's conjecture, stating that there exist at most three. Here we tackle Zauner's conjecture numerically through the construction of Bell inequalities for every pair of integers $n,d \ge 2$ that can be maximally violated in dimension $d$ if and only if $n$ MUBs exist in that dimension. Hence we turn Zauner's conjecture into an optimisation problem, which we address by means of three numerical methods: see-saw optimisation, non-linear semidefinite programming and Monte Carlo techniques. All three methods correctly identify the known cases in low dimensions and all suggest that there do not exist four mutually unbiased bases in dimension six, with all finding the same bases that numerically optimise the corresponding Bell inequality. Moreover, these numerical optimisers appear to coincide with the ``four most distant bases'' in dimension six, found through numerically optimising a distance measure in [P.~Raynal, X.~L\"u, B.-G.~Englert, \textit{Phys.~Rev.~A}, {\bf 83} 062303 (2011)]. Finally, the Monte Carlo results suggest that at most three MUBs exist in dimension ten.

\end{abstract}

\maketitle

\section{Introduction}

Mutually unbiased bases (MUBs), on the one hand, are highly symmetric bases in complex Hilbert spaces and, on the other hand, correspond to pairs of quantum measurements. The defining property of a pair of MUBs is that the overlaps between any two vectors from the two different bases is uniform. This property translates to the corresponding measurements as follows: If a measurement yields a definite outcome when measured on a quantum state, then a measurement unbiased to it will yield a uniformly random outcome on the same state. This feature makes MUBs widely useful in quantum information processing. MUBs were originally introduced in the context of optimal state determination \cite{Ivo81}, but since have been found to be useful in a variety of quantum information processing tasks, such as quantum cryptography \cite{BB84,E91,B98}, quantum communication tasks \cite{THMB15,FK19}, Bell inequalities \cite{BPG03,KSTB+19,TFRB+21} and so on (for a review, see Ref.~\cite{DEBZ10}).

While MUBs have been extensively studied both in the quantum information and the mathematics community for decades, there are still open questions regarding their structure. Most notably, the maximal number of bases that are pairwise mutually unbiased is unknown for general Hilbert space dimension. A general upper bound was shown by Wootters and Fields, stating that in dimension $d$ there exist no more than $d+1$ MUBs~\cite{WF89}. In the same work, they showed that this upper bound is saturated in prime power dimensions, by providing an explicit construction. However, in composite dimensions the only known generic lower bound on the number of MUBs is $p^r+1$, where $p^r$ is the smallest prime power in the prime decomposition of the dimension (this lower bound is shown using tensor products of the Wootters--Fields construction). While in certain dimensions this lower bound has been improved \cite{BW05}, there exists no composite dimension in which the exact number of MUBs is known. For the smallest composite dimension, six, the number of MUBs is known to be no more than seven and no less than three, from the general bounds. However, which of the numbers in between is the exact number of MUBs in dimension six is unknown (apart from the fact that it cannot be six, following from a general result by Weiner \cite{Wei13}). It was first conjectured by Zauner in 1999 that there are no more than three MUBs in dimension six \cite{Zau99}, and this conjecture has not been resolved to date, despite substantial efforts.

There are numerous works trying to prove (or disprove) Zauner's conjecture, both analytically and numerically. While not providing an exhaustive list of references here, let us note that on the analytic side, it has been shown that Zauner's conjecture is equivalent to a conjecture on orthogonal decompositions of Lie algebras \cite{BSTW07}. Furthermore, there exist various analytic constructions of MUB triplet families (see Refs.~\cite{BW09,JMMSzW09} and references therein), but thus far there has not been found even a single vector that is unbiased to all the vectors in any of these triplets. For notable recent developments on Zauner's conjecture see Refs.~\cite{MST21,GP21}.

On the numerical side, Bengtsson et al.~introduced a distance measure of two bases that is maximised if and only if the bases are mutually unbiased \cite{BBEL+07}. This construction turns the problem of finding a set of MUBs into an optimisation problem, maximising all the pairwise distances within a set of bases. Using this approach, Raynal, L\"u and Englert later constructed a two-parameter family of four bases in dimension six, such that for certain values of the parameters, the bases coincide with the numerical maximiser of the distance function \cite{RLE11}. These four bases are not MUBs, but based on the numerical evidence the authors refer to them as ``the four most distant bases in dimension six''.

Since MUBs optimise various quantum information processing tasks, it is natural to measure the closeness of a set of bases to MUBs in terms of some quantum information processing protocol. This was studied formally by Aguilar et al.~\cite{ABMP18}, using the fact that MUBs optimise the success probability of a communication task called quantum random access codes (QRACs). A slight generalisation of the QRAC task is then optimised by a set of $n$ MUBs, and finding $n$ MUBs in dimension $d$ corresponds to optimising the associated success probability. With this method, Aguilar et al.~managed to re-prove the non-existence of $d+2$ MUBs in certain low dimensions using quantum information theoretic tools. However, the case of dimension six remains open.

In this work, we employ similar ideas to tackle Zauner's conjecture. Namely, we study a recently introduced family of Bell inequalities, known to be maximally violated by a pair of MUBs in dimension $d$ \cite{TFRB+21}. We then extend these inequalities to new ones, maximally violated by a set of $n$ MUBs in dimension $d$. Then, we apply three numerical methods for finding the maximal value of these Bell inequalities in a fixed dimension. Namely, we apply see-saw semidefinite programming (SDP), non-linear SDP, and Monte Carlo techniques. While these methods are heuristic---in the sense that there is no guarantee for finding a global maximum---they find the maximum in all the cases where the maximum is known (i.e., it is known that $n$ MUBs exist in the given dimension $d$). Furthermore, when applying these techniques to dimension six and four bases, all the different numerical tools converge to the same bases, and these four bases are---numerically---very close to the ``four most distant bases'' of Ref.~\cite{RLE11} (one should not expect exact equality, since the measure optimised in Ref.~\cite{RLE11} is different from the Bell inequalities we optimise). Hence, our results support Zauner's conjecture, even though no rigorous claim can be made due to the heuristic nature of our methods. Finally, we were able to implement the Monte Carlo algorithm for $d=10$, where---similarly to $d=6$---we do not find more than three MUBs.
%Furthermore, the efficiency of some of our numerical techniques allows us to tackle the number of MUBs in larger composite dimensions, such as 10 and 12.

\section{Preliminaries}

In this section, we introduce the mathematical background and concepts necessary for turning the MUB problem into an optimisation problem. Namely, we formally introduce MUBs, Bell inequalities and the specific family of Bell inequalities tailored for MUBs.

\subsection{Mutually unbiased bases}

Let us take a $d$-dimensional Hilbert space $\mathcal{H} \cong \mathbb{C}^d$, and two orthonormal bases on it, $\{ \ket{ b^1_j } \}_{j=1}^d$ and $\{ \ket{ b^2_k } \}_{k=1}^d$. We say that these two bases are \textit{mutually unbiased} if
\begin{equation}\label{eq:MUB}
  |\braket{ b^1_j }{ b^2_k } |^2 = \frac1d ~~~~ \forall j,k \in [d],
\end{equation}
where $[d] \equiv \{1, 2, \ldots, d\}$. A simple example in dimension two is the computational basis $\{ \ket{0}, \ket{1} \}$ and the Hadamard basis $\{ \frac{1}{\sqrt{2}}( \ket{0} + \ket{1} ), \frac{1}{\sqrt{2}}( \ket{0} - \ket{1} ) \}$.

One may also associate an orthonormal basis with a quantum measurement. In general, a quantum measurement is described by a positive operator-valued measure (POVM), which, in the $d$-dimensional, $d$-outcome case corresponds to a set of $d$ positive semidefinite operators $B_j \ge 0$ on $\mathbb{C}^d$, adding up to the identity operator $\I$. Given an orthonormal basis $\{ \ket{ b_j } \}_{j=1}^d$ on $\mathbb{C}^d$, one can define the corresponding POVM $\{ B_j = \ketbraq{b_j} \}_{j=1}^d$, consisting of rank-1 projections onto the basis elements. We say that two measurements are MUBs if they correspond to a pair of orthonormal bases that are MUBs.

\subsection{Bell inequalities}

We look at MUBs in the context of \textit{Bell scenarios} (see Ref.~\cite{BCPS+14} for a review). Bell scenarios describe physical experiments performed by two distant parties, usually referred to as Alice and Bob. These parties share many copies of a bipartite (quantum) state, and perform local measurements on these copies. The experiment is described by the \textit{correlation}, $p$, with elements $p(a,b|x,y)$
specifying the probability of Alice (Bob) observing locally the outcome $a$ ($b$) upon choosing the measurement setting $x$ ($y$). In quantum theory, the shared state is described by a density operator $\rho \ge 0$ with unit trace ($\tr \rho = 1$) on a tensor product Hilbert space $\mathcal{H}_A \otimes \mathcal{H}_B$. The measurements are described by local POVMs $\{A^x_a\}$ and $\{B^y_b\}$ on the Hilbert spaces $\mathcal{H}_A$ and $\mathcal{H}_B$, respectively. For a fixed state and measurements, the correlation is given by the Born rule,
\begin{equation}\label{eq:correlation}
p(a,b|x,y) = \tr[ \rho ( A^x_a \otimes B^y_b ) ].
\end{equation}
Note that for a pure state, $\rho = \ketbraq{\psi}$, with $\ket{\psi} \in \mathcal{H}_A \otimes \mathcal{H}_B$ and $\braketq{\psi} = 1$, the Born rule reduces to
\begin{equation}\label{eq:correlation_pure}
p(a,b|x,y) = \bra{\psi} A^x_a \otimes B^y_b \ket{\psi}.
\end{equation}

\textit{Bell functionals} are linear functionals of correlations, i.e., functionals of the form
\begin{equation}\label{eq:Bell_ineq}
W(p) = \sum_{a,b,x,y} c_{abxy} p(a,b|x,y),
\end{equation}
where $c_{abxy}$ are real coefficients. Non-trivial \textit{Bell inequalities} are Bell functionals for which $W(p) \le \beta_L$ holds for every correlation of the form
\begin{equation}\label{eq:local}
p(a,b|x,y) = \int_\Lambda \mathrm{d}\mu(\lambda) p_A(a|x,\lambda) p_B(b|y,\lambda)
\end{equation}
(also called a \textit{local} correlation, considered as the notion of classicality in Bell scenarios), but for which there exists a quantum correlation $p$ of the form \eqref{eq:correlation} such that $W(p) > \beta_L$. In Eq.~\eqref{eq:local}, $\Lambda$ is a measurable set with a probability measure $\mu$, $\lambda \in \Lambda$, and $p_A$ and $p_B$ are conditional probability distributions.

While the original interest in Bell inequalities was precisely this separation of local and quantum correlations, we will be interested in their quantum maximum (or \textit{maximal quantum violation}), i.e., the tight upper bound $W(p) \le \beta_Q$ satisfied by all correlations of the form \eqref{eq:correlation}.

\subsection{Bell inequalities for mutually unbiased bases}

In this work, we are interested in a family of Bell inequalities that was introduced in Ref.~\cite{TFRB+21}, and is parametrised by an integer $d \ge 2$. For a fixed $d$, Alice has $d^2$ measurement settings labelled as $x = x_1x_2$ with $x_1, x_2 \in [d]$. Each of these measurements has three outcomes, $a \in \{1, 2, \perp \}$. Bob, on the other hand, has two measurement settings, $y \in \{1,2\}$, with $d$ outcomes each, $b \in [d]$. The Bell inequality then reads
\begin{equation}\label{eq:Bell2MUB}
\begin{split}
W_d(p) & \left. = \sum_{x_1,x_2,y} \big[ p( y, x_y | x_1x_2, y ) - p( \bar{y}, x_y | x_1x_2, y) \big] \right. \\
& \left. - \frac12 \sqrt{ \frac{d-1}{d} } \sum_{x_1,x_2} \big[ p_A( 1 | x_1x_2 ) + p_A( 2 | x_1x_2 ) \big], \right.
\end{split}
\end{equation}
where $\bar{y}$ flips the value of $y \in \{1,2\}$, and $p_A(a|x) = \sum_b p(a,b|x,y)$ is the marginal probability distribution of Alice (which is independent of $y$).

This is a non-trivial Bell inequality with maximal quantum violation $\beta_Q = \sqrt{d(d-1)}$ \cite{TFRB+21}. The maximal violation can be achieved with the maximally entangled $d$-dimensional state $\ket{ \phi^+_d } \equiv \frac{1}{\sqrt{d}} \sum_{j=1}^d \ket{j} \otimes \ket{j}$, and \textit{any} pair of MUB measurements on Bob's side. Moreover, if the dimension is fixed to be $d$, this is the only way in which the maximal quantum violation can be achieved, up to local unitary freedom \cite{TFRB+21}. This property of the Bell inequality \eqref{eq:Bell2MUB} forms the core of our numerical approaches to construct MUBs. 

The above Bell inequality can be straightforwardly extended to a set of $n$ measurements on Bob's side. The Bell inequality for $n$ measurements is a sum of Bell inequalities of the form \eqref{eq:Bell2MUB}. For each pair $y,z \in [n]$ such that $y<z$ (denoted in the following as $(y,z) \in \text{Pairs}[n]$), we introduce $d^2$ settings for Alice, labelled as $x=(y,z)x_y x_z$ with $x_y, x_z \in [d]$, and take a copy of the Bell inequality in Eq.~\eqref{eq:Bell2MUB}, defined as
\begin{equation}\label{eq:Bellyz}
\begin{split}
W^{(y,z)}_d(p) & \left. = \sum_{x_y,x_z,w} \big[ p( a_w , x_w | (y,z)x_y x_z, w ) \right. \\
& \left. - p( \bar{a}_w, x_w | (y,z) x_y x_z, w) \big] \right. \\
& \left. - \frac12 \sqrt{ \frac{d-1}{d} } \sum_{x_y,x_z} \big[ p_A( 1 | (y,z) x_y x_z ) \right. \\
& \left. + p_A( 2 | (y,z) x_y x_z ) \big], \right.
\end{split}
\end{equation}
where $w \in \{y, z\}$, $a_y = 1$, $a_z = 2$, and $\bar{a}_w$ flips the value of $a_w$. The final Bell inequality %for $n$ $d$-dimensional MUBs 
then reads
\begin{equation}\label{eq:BellnMUB}
W_{d,n}(p) = \sum_{(y,z) \in \text{Pairs}[n]} W^{(y,z)}_d(p).
\end{equation}
It is clear that $W_{d,n}(p) \le \binom{n}{2} \sqrt{d(d-1)}$, by applying the known bound to each individual term in the above sum. Moreover, if the dimension is $d$, the only way to reach this maximum (up to local unitary freedom) is by using the maximally entangled state, and if the $n$ measurements on Bob's side correspond to MUBs. Hence, we can reformulate the MUB problem in terms of these Bell inequalities:
\begin{prop}\label{prop:BellMUB}
$W_{d,n}(p) = \binom{n}{2} \sqrt{d(d-1)}$ can be achieved in dimension $d$ if and only if $n$ MUBs exist in dimension $d$.
\end{prop}

\subsection{The optimisation problem}

According to the above proposition, finding $n$ MUBs in dimension $d$ can be cast as an optimisation problem, maximising $W_{d,n}(p)$ over $d$-dimensional quantum states and measurements. To see how to do this explicitly, let us first write out the Bell inequality $W_d$ in Eq.~\eqref{eq:Bell2MUB} in terms of a quantum state $\ket{ \psi }$ and measurements $\{A^x_a\}$, $\{B^y_b\}$, using Born's rule in Eq.~\eqref{eq:correlation_pure}:
\begin{equation}\label{eq:Bell2MUBQ}
\begin{split}
W_d( \ket{\psi}, & \left. \{A^x_a\}, \{B^y_b\} ) = \sum_{x_1,x_2,y} \big( \bra{ \psi } A^{x_1 x_2}_y \otimes B^y_{x_y} \ket{\psi} \right. \\
& \left. - \bra{ \psi } A^{x_1 x_2}_{ \bar{y} } \otimes B^y_{x_y} \ket{\psi} \big) \right. \\
& \left. - \frac12 \sqrt{ \frac{d-1}{d} } \sum_{x_1,x_2} \big( \bra{ \psi } A^{x_1 x_2}_1 \otimes \I \ket{\psi} \right. \\
& \left. + \bra{ \psi } A^{x_1 x_2}_2 \otimes \I \ket{\psi} \big). \right. \\
& \left.  = \sum_{j,k} \bra{ \psi } \big[ ( A^{jk}_1 - A^{jk}_2 ) \otimes (B^1_{j} - B^2_{k} ) \right. \\
& \left. - \frac12 \sqrt{ \frac{d-1}{d} } ( A^{jk}_1 + A^{jk}_2 ) \otimes \I \big] \ket{ \psi }, \right.
\end{split}
\end{equation}
where in the second equality we wrote out the summation over $y$ and switched to the notation $x_1 x_2 \to jk$ with $j,k \in [d]$. Maximising the inequality \eqref{eq:Bell2MUBQ} in terms of the state and the measurements can then be written as the optimisation problem
\begin{equation}
\begin{split}
\max_{ \ket{\psi}, \{A^x_a\}, \{B^y_b\} } ~~~ & \left. W_d( \ket{\psi}, \{A^x_a\}, \{B^y_b\} ) \right. \\
\text{s.t.} \quad \quad \quad & \left. \ket{\psi} \in \mathbb{C}^d \otimes \mathbb{C}^d, ~~ \braketq{\psi} = 1 \right. \\
& \left. A^x_a, B^y_b \in \mathcal{L}_{\text{sa}}( \mathbb{C}^d ) ~~ \forall a,b,x,y \right. \\
& \left. A^x_a \ge 0 ~~ \forall a,x, ~~ B^y_b \ge 0 ~~ \forall b,y \right. \\
& \left. \sum_a A^x_a = \I ~~ \forall x, ~~ \sum_b B^y_b = \I ~~ \forall y, \right.
\end{split}
\end{equation}
where $\mathcal{L}_{\text{sa}}( \mathbb{C}^d )$ is the set of self-adjoint linear operators on $\mathbb{C}^d$. Optimising $W_{d,n}$ can be written in a similar fashion, with:
\begin{equation}\label{eq:BellnMUBQ}
\begin{split}
W_{d,n} & \left. ( \ket{\psi}, \{A^x_a\}, \{B^y_b\} ) = \sum_{(y,z) \in \text{Pairs}[n]} \Bigg\{ \right. \\
& \left. \sum_{j,k} \bra{ \psi } \big[ ( A^{(y,z) jk}_1 - A^{(y,z) jk}_2 ) \otimes (B^y_{j} - B^z_{k} ) \right. \\
& \left. - \frac12 \sqrt{ \frac{d-1}{d} } ( A^{(y,z) jk}_1 + A^{(y,z) jk}_2 ) \otimes \I \big] \ket{ \psi } \Bigg\}. \right.
\end{split}
\end{equation}
The optimisation problem is then
\begin{equation}\label{eq:optimisation_general}
\begin{split}
\max_{ \ket{\psi}, \{A^x_a\}, \{B^y_b\} } ~~~ & \left. W_{d,n}( \ket{\psi}, \{A^x_a\}, \{B^y_b\} ) \right. \\
\text{s.t.} \quad \quad \quad & \left. \ket{\psi} \in \mathbb{C}^d \otimes \mathbb{C}^d, ~~ \braketq{\psi} = 1 \right. \\
& \left. A^x_a, B^y_b \in \mathcal{L}_{\text{sa}}( \mathbb{C}^d ) ~~ \forall a,b,x,y \right. \\
& \left. A^x_a \ge 0 ~~ \forall a,x, ~~ B^y_b \ge 0 ~~ \forall b,y \right. \\
& \left. \sum_a A^x_a = \I ~~ \forall x, ~~ \sum_b B^y_b = \I ~~ \forall y. \right.
\end{split}
\end{equation}

In this work, we consider various approaches to solve the optimisation problem \eqref{eq:optimisation_general}. In particular, from Proposition~\ref{prop:BellMUB} it follows that $n$ MUBs exist in dimension $d$ if and only if the solution of the above optimisation problem is $\binom{n}{2}\sqrt{d(d-1)}$. We will facilitate the problem using knowledge about the optimal realisation of the Bell inequality from Ref.~\cite{TFRB+21}.

First of all, we notice that the value $\binom{n}{2}\sqrt{d(d-1)}$ can only be achieved in dimension $d$ with the maximally entangled state~\cite{TFRB+21}. Without loss of generality, we therefore impose that $\ket{\psi} = \ket{\phi^+_d} = \frac{1}{\sqrt{d}} \sum_{j=1}^d \ket{j} \otimes \ket{j}$. We can then use the fact that for any two operators $A$ and $B$ on $\mathbb{C}^d$ we have that $\bra{ \phi^+_d } A \otimes B \ket{ \phi^+_d } = \frac1d \tr( A^T B )$, where $(.)^T$ is the transposition in the basis $\{ \ket{j} \}$. As a second simplification, we notice that in order to saturate the bound $\binom{n}{2}\sqrt{d(d-1)}$ in dimension $d$, Alice's measurement operators $A^x_1$ and $A^x_2$, and all of Bob's measurement operators must be trace-1 \cite{TFRB+21}. For such operators we have that $\bra{ \phi^+_d } A^x_1 \otimes \I \ket{ \phi^+_d } = \bra{ \phi^+_d } A^x_2 \otimes \I \ket{ \phi^+_d } = \frac1d$. The second term in Eq.~\eqref{eq:BellnMUBQ} is then a constant, $- \binom{n}{2} \sqrt{d(d-1)}$, and does not influence the optimisation problem. The simplified Bell expression finally reads
\begin{equation}\label{eq:BellnMUB+}
\begin{split}
W^+_{d,n} & \left. ( \{A^x_a\}, \{B^y_b\} ) = \sum_{(y,z) \in \text{Pairs}[n]} \Bigg\{ \right. \\
& \left. \frac1d \sum_{j,k} \tr \big( (A^{(y,z) jk}_1 - A^{(y,z) jk}_2)^T (B^y_{j} - B^z_{k}) \big) \Bigg\}, \right.
\end{split}
\end{equation}
and its maximum quantum value $W^+_{d,n}$ satisfies:
\begin{equation}
    W^+_{d,n} \le W_{\text{MUB}}(d,n) \equiv n(n-1)\sqrt{d(d-1)} ~.
\end{equation}
The simplified optimisation problem becomes
\begin{equation}\label{eq:optimisation_+}
\begin{split}
\max_{ \{A^x_a\}, \{B^y_b\} } ~~~ & \left. W^+_{d,n}( \{A^x_a\}, \{B^y_b\} ) \right. \\
\text{s.t.} \quad \quad & \left.  A^x_a, B^y_b \in \mathcal{L}_{\text{sa}}( \mathbb{C}^d ) ~~ \forall a,b,x,y \right. \\
& \left. A^x_a \ge 0 ~~ \forall a,x, ~~ B^y_b \ge 0 ~~ \forall b,y \right. \\
& \left. \tr A^x_a = 1 ~~ \forall a,x, ~~ \tr B^y_b = 1 ~~ \forall y,b \right. \\
& \left. \sum_a A^x_a = \I ~~ \forall x, ~~ \sum_b B^y_b = \I ~~ \forall y. \right.
\end{split}
\end{equation}
The optimal value of this optimisation problem is $n(n-1)\sqrt{d(d-1)}$ if and only if $n$ MUBs exist in dimension~$d$.

We may further simplify the optimisation problem.
%, using properties and relations of the measurement operators.
From Ref.~\cite{TFRB+21} we know that in the optimal realisation the $B^y_b$ operators are rank-1 projections, $B^y_j = \ketbraq{ b^y_j }$ (they are projections onto the basis elements of MUBs). For such operators, we have that
\begin{equation}
(B^y_j - B^z_k)^3 = [ 1 - \tr( B^y_j B^z_k ) ] (B^y_j - B^z_k),
\end{equation}
which implies that the spectrum of $B^y_j - B^z_k$ is contained in $\{0, \pm \lambda^{yz}_{jk} \}$, where $\lambda^{yz}_{jk} \equiv \sqrt{ 1 - \tr( B^y_j B^z_k ) } = \sqrt{ 1 - |\braket{ b^y_j }{ b^z_k }|^2 }$. Moreover, we have that $\tr[ (B^y_j - B^z_k)^2 ] = 2 (\lambda^{yz}_{jk})^2$ and $\tr(B^y_j - B^z_k) = 0$, and therefore $B^y_j - B^z_k$ has one eigenvalue $\lambda^{yz}_{jk}$, one eigenvalue $-\lambda^{yz}_{jk}$ and the rest of the eigenvalues are~0.
Furthermore, in the optimal realisation we have that $(A^{(y,z)jk}_1)^T$ is the rank-1 projection onto the eigenspace of $B^y_j - B^z_k$ corresponding to $\lambda^{yz}_{jk}$, and $(A^{(y,z)jk}_2)^T$ is the rank-1 projection onto the eigenspace of $B^y_j - B^z_k$ corresponding to $-\lambda^{yz}_{jk}$. With these final simplifications the Bell expression reads
\begin{eqnarray}
W^{+B}_{d,n} ( \{B^y_b\} ) &=& \frac2d \sum_{(y,z) \in \text{Pairs}[n]} \sum_{j,k} \sqrt{ 1 - \tr( B^y_j B^z_k ) } \label{eq:BellnMUB+B}\\
&=& \frac2d \sum_{(y,z) \in \text{Pairs}[n]} \sum_{j,k} \sqrt{ 1 - |\langle b_j^y | b_k^z \rangle|^2} \label{eq:MUBness_measure}
\end{eqnarray}
and the corresponding optimisation problem is
\begin{equation}\label{eq:optimisation_+B}
\begin{split}
\max_{ \{B^y_b\} } ~~~ & \left. W^{+B}_{d,n}( \{B^y_b\} ) \right. \\
\text{s.t.} \quad & \left.  B^y_b \in \mathcal{L}_{\text{sa}}( \mathbb{C}^d ) ~~ \forall a,b,x,y \right. \\
& \left. (B^y_b)^2 = B^y_b ~~ \forall b,y \right. \\
& \left. \tr B^y_b = 1 ~~ \forall b,y \right. \\
& \left. \sum_b B^y_b = \I ~~ \forall y. \right.
\end{split}
\end{equation}
Note that projectivity and self-adjointness implies positive semidefiniteness, and therefore positive semidefiniteness does not need to be imposed.

The optimal value of the optimisation problem \eqref{eq:optimisation_+B}, denoted as $W_{\rm max}(d,n)$, is $n(n-1)\sqrt{d(d-1)}$ if and only if $n$ MUBs exist in dimension~$d$.
In fact, Alice's measurements have been completely removed from the problem, and one may regard Eq.~\eqref{eq:optimisation_+B} as a purely geometrical problem: find $n$ orthonormal bases, $\big\{ \{ \ket{b_j^y} ~|~ j\in [d] \} ~|~ y\in[n] \big\}$, maximising the ``MUB-ness measure'' of Eq.~\eqref{eq:MUBness_measure}.

In the following three sections we apply three numerical methods to solve the optimisation problems \eqref{eq:optimisation_+} and \eqref{eq:optimisation_+B} in order to numerically tackle Zauner's conjecture.

\section{See-saw SDP}

\subsection{Methodology}

One arrives at a relatively simple method of optimising problem \eqref{eq:optimisation_+} by noticing that the objective function is bi-linear in the $A^x_a$ and $B^y_b$ matrices. That is, if every $A^x_a$ is fixed, then the problem simplifies to optimising a linear functional of the $B^y_b$ matrices with a series of linear and positive semidefinite constraints. This is a standard SDP, for which there exist efficient solvers.

The \emph{see-saw} optimisation technique starts with fixing the set of $A^x_a$ matrices satisfying the constraints of the problem \eqref{eq:optimisation_+}, either with random values or based on some prior knowledge. We then solve the problem for the $B^y_b$ matrices, which is a standard SDP. Then, we fix the $B^y_b$ matrices to the optimum found, and solve the resulting SDP for the $A^x_a$ matrices, and so on. By repeating this process, the system eventually converges to a stable result, i.e.~the value of the objective function does not change beyond a given threshold  within a certain window of iterations (a change of less than $10^{-9}$ for $10$ iterations in our implementation). Although the see-saw method has never been proven to  converge to the global optimum, for our current problem it has never failed to converge within the chosen precision if given sufficient time, always to the value expected (i.e., to $W_{\text{MUB}}(d,n)$ whenever it is known that $n$ MUBs exist in dimension $d$). We implemented the see-saw algorithm with the help of the SDP solving library MOSEK \cite{mosek}.

\subsection{Results}

The values obtained with the see-saw algorithm are shown in Table \ref{tbl:seesawResults} (the values displayed are $1-W_{d,n}/W_\text{MUB}(d,n)$ for easier comparison across different $n$ and $d$, where $W_{d,n}$ is the result of the optimisation). Notice that whenever $n$ MUBs exist in dimension $d$, the see-saw method correctly converges to the MUB solution, and whenever it is known that $n$ MUBs do not exist in dimension $d$, the method does indeed converge to a value less than $W_\text{MUB}(d,n)$. For the unknown case of four MUBs in dimension six, the see-saw method could not find four MUBs, supporting Zauner's conjecture. Furthermore, the optimal measurements found by the see-saw method are numerically very close to the ``four most distant bases'' of Ref.~\cite{RLE11} (see also Section \ref{subsec:d6n4}). Note, however, that the results simply mean that the method could not find four MUBs in dimension six, but one cannot rule out the possibility that they exist.

The method also has consistent convergence, for example, for $d=2$ and $n=2$, all of $10000$ see-saw optimisations from random starts converged to the correct value of $W_\text{MUB}(2,2) = 2.82843$ up to five decimal places, albeit finding a different set of optimum matrices. Similarly, for $10000$ optimisations for $d=2$ and $n=4$, all optimisations converged to the same value of $16.72616$ (correctly signifying non-existence). We did not perform a similar sized convergence analysis for the larger dimensions due to the time required to optimise, however, several runs were always performed to certify a minimal level of consistency.

All of the results for this method were obtained on a desktop PC with 8GB of RAM using 4 cores, with times varying between milliseconds for the smallest problem ($d=2$, $n=2$) and hours for the largest one ($d=6$, $n=4$). Since SDP solvers are efficiently parallelised, this method offers good parallel scaling, however, the memory requirement is the highest of all of our methods, since it requires explicit storage and optimisation of the $A^x_a$ matrices.

% \begin{table*}[t]
%     \centering
%     \begin{tabular}{ | c | c c c c c | } 
%         \hline
%         \backslashbox{n}{d} & 2 & 3 & 4 & 5 & 6 \\ 
%         \hline
%         2 & \textcolor{blue}{2.82843} & \textcolor{blue}{7.34847} & \textcolor{blue}{13.85641} & \textcolor{blue}{22.36068} & \textcolor{blue}{32.86335} \\ 
%         3 & \textcolor{blue}{8.48528} & \textcolor{blue}{22.04541} & \textcolor{blue}{41.56922} & \textcolor{blue}{67.08204} & \textcolor{blue}{98.59005} \\ 
%         4 & \textcolor{red}{16.72616} & \textcolor{blue}{44.09081} & \textcolor{blue}{83.13844} & \textcolor{blue}{134.16407} & \textcolor{red}{197.17175} \\ 
%         5 & \textcolor{red}{-} & \textcolor{red}{73.19725} & \textcolor{blue}{138.56405} & \textcolor{blue}{223.60678} & \textcolor{red}{-} \\ 
%         6 & \textcolor{red}{-} & \textcolor{red}{-} & \textcolor{red}{207.46035} & \textcolor{blue}{335.41017} & \textcolor{red}{-} \\ 
%         7 & \textcolor{red}{-} & \textcolor{red}{-} & \textcolor{red}{-} & \textcolor{red}{469.14817} & \textcolor{red}{-} \\
%         \hline
%     \end{tabular}
%     \caption{The value obtained with the seesaw method after convergence for various dimensions ($d$) and number of bases ($n$). Blue text indicates that the value matched the known maximum, thus there do exist MUBs for that problem (within this precision). The red indicates that the value was at least $10^{-5}$ away from the known maximum, thus it is unlikely that MUBs exist for that problem.}
%     \label{tbl:seesawResults}
% \end{table*}

\begin{table*}[t]
    \centering
    \begin{tabular}{ | c | c c c c c | } 
        \hline
        \backslashbox{$n$}{$d$} & 2 & 3 & 4 & 5 & 6 \\ 
        \hline
        2 & 0.00000 & 0.00000 & 0.00000 & 0.00000 & 0.00000 \\ 
        3 & 0.00000 & 0.00000 & 0.00000 & 0.00000 & 0.00000 \\ 
        4 & \textbf{0.01440} & 0.00000 & 0.00000 & 0.00000 & \textbf{0.00004} \\ 
        5 & - & \textbf{0.00391} & 0.00000 & 0.00000 & - \\ 
        6 & - & - & \textbf{0.00186} & 0.00000 & - \\ 
        7 & - & - & - & \textbf{0.00091} & - \\
        \hline
    \end{tabular}
    \caption{The values $1-W_{d,n}/W_\text{MUB}(d,n)$, where $W_{d,n}$ is the result of the optimisation, obtained with the see-saw method after convergence for various dimensions ($d$) and numbers of bases ($n$), to 5 decimal places. The values depicted are consistently obtained by multiple runs. The values in bold indicate that the value was at least $10^{-5}$ away from zero, meaning that the method could not find $n$ MUBs in dimension $d$. The ``-'' symbol indicates that we have not performed the optimisation. Note that in all the known cases, the algorithm predicts correctly existence/nonexistence, and it predicts that four MUBs do not exist in dimension six.}
    \label{tbl:seesawResults}
\end{table*}

\section{Non-linear SDP}

\subsection{Methodology}

An alternative approach to the optimisation is to focus on problem \eqref{eq:optimisation_+B}, which features only the $B^y_b$ matrices and thus contains fewer variables for a reduction in the size of the search space as well as the memory required. The downside, however, is that the objective function is now non-linear (not even bi-linear) and thus many efficient solvers (i.e.~for standard SDP systems) can no longer be applied. To optimise this problem we adapt a method based on the work by Yamashita et al.~for optimising a non-linear SDP using a primal-dual interior point method \cite{yamashita2012}. This method, assuming a few basic conditions (discussed later), is guaranteed to converge to a Karush–Kuhn–Tucker (KKT) point, a point satisfying a series of constraints known as the KKT conditions, which are necessary for optimality \cite{BV04}.

These KKT conditions are only sufficient (imply a global minimum) in a subset of cases, the main of which being that the problem is convex, which is unfortunately not true in our case. An interesting property of our search-space, however, is that by taking the derivative of our objective function it can be shown that all local minima are global minima if the constraints are met, thus implying that our search space is, in fact, a series of discontinuous convex regions. This signifies that for the MUB-existence case this method will always converge to MUBs, although no claim can be made for the non-existence case.

In order to implement the method of Yamashita et al., we parametrise the measurement operators $\{B^x_y\}$ by a real vector $\bx = (x_i)_i$. To be able to deal with real numbers instead of complex ones, we note that every self-adjoint matrix $B = B_r + \ii B_i$ (where $B_r$ is real symmetric and $B_i$ is real anti-symmetric) can be mapped to the real symmetric matrix $\hat{B}$ via
\begin{equation}\label{eq:realB}
    B \mapsto \hat{B} =
    \begin{bmatrix}
        B_r & B_i \\
        -B_i & B_r
    \end{bmatrix}.
\end{equation}
It is easy to verify that $B \ge 0$ if and only if $\hat{B} \ge 0$, and $\tr B = \frac12 \tr \hat{B}$.

We therefore define---in line with the method of Yamasitha et al.---a matrix $X(\bx) = \sum_i C_i x_i + D$, which is a block diagonal matrix containing the $\hat{B}^y_b$ matrices on its diagonal in such a way that the linear constraints $\sum_b B^y_b = \I$ and $\tr B^y_b = 1$ of the optimisation problem \eqref{eq:optimisation_+B} are already enforced. The real parameters $x_i$ correspond to those elements of the $\hat{B}^y_b$ matrices that are free after enforcing the linear constraints. While the constraint $B^y_b \ge 0$ is superfluous for the problem \eqref{eq:optimisation_+B}, we chose to include this in our optimisation problem, as this constraint is heavily used in the method of Yamashita et al. With the parametrisation above, this constraint is equivalent to $X(\bx) \ge 0$. The last remaining constraint is projectivity, $(B^y_b)^2 = B^y_b$ for all $y$ and $b$, which is equivalent to $X^2(\bx) = X(\bx)$. We enforce this constraint through $g(\bx) \equiv |\!|X^2(\bx) - X(\bx)|\!|_F^2 = 0$, where $|\!|.|\!|_F$ is the Frobenius norm. Further, we denote the objective function in terms of $\bx$ by $W(\bx)$, suppressing the $d,n$ indices whenever it does not lead to confusion.

The method requires introducing Lagrange multipliers (dual variables) for every constraint. In our case, there is a single inequality constraint $X(\bx) \ge 0$, to which we assign the dual variable $Z$, which is a matrix with the same dimensions as $X(\bx)$. Furthermore, we have a single equality constraint, $g(\bx) = 0$, to which we assign the dual variable $y$, which is a scalar. The resulting Lagrangian reads
\begin{equation}\label{eqn:kkt_lagrangian}
    L(\bx,y,Z) = W(\bx) - y g(\bx) - \tr[ Z^T X(\bx) ].
\end{equation}

The algorithm for solving the optimisation problem is iterative, and each iteration begins with the calculation of $G$, the Hessian of the Lagrangian. In our case this can be quite expensive so we opt to use the alternative update method also proposed in Ref.~\cite{yamashita2012} based on the Broyden–Fletcher–Goldfarb–Shanno (BFGS) algorithm, which approximates the Hessian without requiring the full calculation of the second derivatives. This $G$ is then used to form a series of linear equations which we solve using a stabilised bi-conjugate gradient method in order to obtain the update directions for the primal ($\bx$) and dual ($y,Z$) variables. Following this, a simple line search is performed to find the optimum step size, then the variables are updated. This process is then repeated until the following barrier KKT conditions are met for some barrier parameter $\mu$:

\begin{align}\label{eqn:kkt_conditions}
    r(\bx, y, Z, \mu) \equiv \begin{pmatrix} \nabla L(\bx,y,Z) \\ g(\bx) \\ X(\bx)Z-\mu I \end{pmatrix} = \begin{pmatrix} 0 \\ 0 \\ 0 \end{pmatrix}
\end{align}

Through repeated iterations of this method with values of $\mu$ converging towards zero, this algorithm has been proven to always converge to a KKT point of the system, assuming certain conditions. The first of these is that the functions $W(\bx)$ and $g(\bx)$ are both twice continuously differentiable, which in our case is true: $W(\bx)$ is simply a sum of square roots of polynomials of $\bx$, whilst $g(\bx)$ is a vector of polynomials of $\bx$. The second condition is that the vector $\bx$ must remain within a finite set during the optimisation, which for us is true since infinite values are non-optimal for the MUB functional. The third condition is that the matrices $C_i$ must be linearly independent, which for us is true since they serve to place a single component of $\bx$ to one (diagonal) or two (off-diagonal) positions in the matrix $X(\bx)$.

Similarly to the the see-saw method, there exists some non-determinism to this method due to it starting from a randomised interior point. This is found by first beginning with a random vector, then performing gradient descent using the derivatives of the constraints until the vector satisfies all constraints within some precision. This vector is then used as the starting point for the iterative method. Although perhaps one could start with a ``good guess'' for the vector and optimise from there, we decided to begin from a random point to attempt to cover a wider search space.

\subsection{Results}

The values obtained through this algorithm are shown in Table \ref{tbl:kktResults}. The optimal values, as well as the optimal bases found, agree with the result of the see-saw method up to high numerical precision. In particular, we find the same set of four bases to numerically optimise the six dimensional case (see Section \ref{subsec:d6n4} for further details).

Regarding performance, a large amount of time can be saved if certain parameters, such as the initial step size, are chosen correctly. Notably, there appears to be no universally optimal set of parameters, and many of the results are obtained through the manual tweaking of the parameters for the specific system. All of the results for this method were obtained on a standard desktop PC using 4 cores, with the times varying between milliseconds for the smallest problem and an hour for the largest one. This method has a significantly reduced memory cost compared to the see-saw method, since here the $A^x_a$ matrices are not used, as well as in general offering faster optimisation.

\begin{table*}[t]
    \centering
    \begin{tabular}{ | c | c c c c c | } 
        \hline
        \backslashbox{$n$}{$d$} & 2 & 3 & 4 & 5 & 6 \\ 
        \hline
        2 & 0.00000 & 0.00000 & 0.00000 & 0.00000 & 0.00000 \\ 
        3 & 0.00000 & 0.00000 & 0.00000 & 0.00000 & 0.00000 \\ 
        4 & \textbf{0.01440} & 0.00000 & 0.00000 & 0.00000 & \textbf{0.00004} \\ 
        5 & - & \textbf{0.00391} & 0.00000 & 0.00000 & - \\ 
        6 & - & - & \textbf{0.00161} & 0.00000 & - \\
        7 & - & - & - & \textbf{0.00091} & - \\
        \hline
    \end{tabular}
    \caption{The values $1-W_{d,n}/W_\text{MUB}(d,n)$, where $W_{d,n}$ is the result of the optimisation, obtained at an approximate KKT point with $|\!|r(\bx,y,Z,\mu)|\!| \equiv \sqrt{ |\!| \nabla L(\bx,y,Z) |\!|^2 + |g(\bx)|^2 + |\!| X(\bx) Z - \mu I |\!|_F^2} \le 10^{-5}$ for various dimensions ($d$) and numbers of bases ($n$), to 5 decimal places. The values depicted are consistently obtained by multiple runs. The values in bold indicate that the value was at least $10^{-5}$ away from zero, meaning that the method could not find $n$ MUBs in dimension $d$. Note that in all the known cases, the algorithm predicts correctly existence/nonexistence, and it predicts that four MUBs do not exist in dimension six.}
    \label{tbl:kktResults}
\end{table*}

% \begin{table*}[t]
%     \centering
%     \begin{tabular}{ | c | c c c c c | } 
%         \hline
%         \backslashbox{n}{d} & 2 & 3 & 4 & 5 & 6 \\ 
%         \hline
%         2 & {9} & {230} & {1747} & \textcolor{black}{14908} & \textcolor{black}{41292} \\ 
%         3 & \textcolor{black}{56} & \textcolor{black}{746} & \textcolor{black}{7512} & \textcolor{black}{39618} & \textcolor{black}{339872} \\ 
%         4 & \textcolor{black}{267} & \textcolor{black}{3426} & \textcolor{black}{21742} & \textcolor{black}{233443} & \textcolor{black}{2713456} \\ 
%         5 & \textcolor{black}{1373} & \textcolor{black}{9699} & \textcolor{black}{19846} & \textcolor{black}{349549} & \textcolor{black}{-} \\ 
%         6 & \textcolor{black}{2883} & \textcolor{black}{13009} & \textcolor{black}{305446} & \textcolor{black}{588605} & \textcolor{black}{-} \\ 
%         \hline
%     \end{tabular}
%     \caption{The time in milliseconds required to obtain a KKT point for various dimensions ($d$) and number of bases ($n$), corresponding to each of the results in table \ref{tbl:kktResults}. Each of these optimisations has parameters tuned for that problem, for instance: if the iterations appeared to fluctuate around the known optimal then the scaling of the step size is reduced, leading to a more consistent descent.}
%     \label{tbl:kktTimes}
% \end{table*}

\section{The MUB problem as a ground state problem}
\begin{table*}
    \centering
    \begin{tabular}{ | c | c c c c c c c c c| } 
        \hline
        %\backslashbox{$n$}{$d$} & 2 & 3 & 4 & 5 & 6 & \hspace{5pt} 7\hspace{5pt} & \hspace{5pt} 8\hspace{5pt} &  \hspace{5pt} 9\hspace{5pt} &  \hspace{5pt} 10 \hspace{5pt} \\ 
                \backslashbox{$n$\kern-0.5em}{\kern-0.5em$d$} & 2 & 3 & 4 & 5 & 6 & 7 & 8 & 9 & 10 \\ 
        \hline
        2 & 0. & 0. & 0. & 0. & 0. & 0. & 0. & 0. & 0.  \\ 
        \hline
        3 & 0. & 0. & 0. & 0. & 0. & 0. & 0. & 0. & 0.  \\ 
        \hline
        4 & \textbf{0.01440} & 0. & 0. & 0. & \textbf{0.00004215} & 0. & 0. & 0. & \textbf{0.000005938}   \\ 
        \hline
        5 & {-} & \textbf{0.003912} & 0. & 0. & {-} & 0.  & 0. & 0. & {-}  \\ 
        \hline
        6 & {-} & {-} & \textbf{0.001609} & 0. & {-}  & 0. & 0. & {-} & {-} \\ 
        \hline
        7 & {-} & {-} & {-} & \textbf{0.0009073} & {-}  & 0. & 0. & {-} & {-}\\ 
        \hline
        8 & {-} & {-} & {-} & {-} & {-}  & 0. & 0. & {-} & {-}\\ 
        \hline
    \end{tabular}
    \caption{Monte Carlo results. Relative deviation from the MUB optimum in Eq.~\eqref{eq_W_MonteCarlo} $1-W_{\rm max} / W_{\rm MUB}$. ``$0.$'' indicates that $n$ MUBs have been found in dimension $d$ up to at least $10^{-10}$ precision. The positive values in bold are the numerical maxima found over at least three independent Monte Carlo simulations. For $d=10$, $n=4$, three out of ten independent simulations found the indicated value; it remains possible that this does not correspond yet to the true optimum.}
    \label{tbl:montecarloResults}
\end{table*}

\subsection{Methodology}
Finally, we apply an optimisation method inspired by statistical physics, namely simulated annealing \cite{simulated_annealing}. Given $n$ orthonormal bases $\{|b_j^y\rangle; j\in [d]\} ({y\in[n]})$ in a Hilbert space of dimension $d$, our goal is to maximise the expression of Eq.~\eqref{eq:MUBness_measure}, which we rewrite here for completeness (we keep the $d,n$ dependence implicit throughout this section, and keep the $+B$ superscript implicit in Eq.~\eqref{eq:MUBness_measure}):
\begin{equation}
    W[{\bf x}] = \frac2d \sum_{(y,z)\in{\rm Pairs}[n]} \sum_{j,k=1}^d \sqrt{1 - |\langle b_j^y | b_k^z \rangle|^2},
    \label{eq_W_MonteCarlo}
\end{equation}
where ${\bf x} = \{|b_j^y\rangle\}$ defines the collection of $n$ bases. When the $n$ bases are mutually unbiased ($|\langle b_j^y|b_k^z\rangle|^2=1/d$ for all $y,z \in \text{Pairs}[n]$ and all $j,k \in [d]$), we have $W = W_{\rm MUB}=n(n-1)\sqrt{d(d-1)}$. In general, the maximal value is $W_{\rm max} \le W_{\rm MUB}$, with equality if and only if $n$ MUBs exist in dimension $d$. Our strategy is then to optimise the bases by maximising $W$ (or equivalently, minimising $-W$) via simulated annealing. As detailed below, this amounts to parametrising the vectors $|b_j^y\rangle$ by some parameters ${\bf x}$, and regarding $-W({\bf x})$ as the energy of the configuration ${\bf x}$. The ground state (namely, the lowest-energy configuration) of $-W$ is found by sampling ${\bf x}$ with probability $\propto \exp[\beta W({\bf x})]$, progressively ramping up the inverse temperature $\beta$. In contrast to gradient-descent-based approaches, this allows to explore a variety of local minima of $-W$ at nonzero temperature, with the hope of finally converging to the global minimum when the temperature approaches zero. As our results show, this hope is indeed confirmed by solid evidence. If $n$ MUBs do exist, they are obtained at the end of the optimisation, saturating the value $W_{\rm max}=W_{\rm MUB}$. Otherwise, the algorithm converges to an optimum (presumably the global optimum) $W_{\rm max} < W_{\rm MUB}$, supporting that $n$ MUBs do not exist in dimension $d$.

{\bf Parametrising the bases.--}
Each vector $|b_j^y\rangle$ is simply parametrised by its decomposition in the canonical basis: $|b_j^y\rangle = \sum_{i=1}^d U_{ji}^y |e_i\rangle$, where ${\bf x}:=\{U^y\}_{y=1}^n$ are $n$ complex $d\times d$ unitary matrices. The overlaps are then obtained as $\langle b_j^y|b_k^z\rangle= \sum_{i=1}^d (U_{ji}^y)^* U_{ki}^z$, where $(.)^\ast$ is the complex conjugation.

{\bf Simulated annealing.--}
The basic idea of simulated annealing \cite{simulated_annealing} is to consider:
\begin{equation}
	\langle W \rangle_\beta = \int d{\bf x} ~W({\bf x}) ~ \frac{e^{\beta W({\bf x})}}{Z_\beta}~,
	\label{eq_thermal_avg}
\end{equation}
 where $Z_\beta = \int d{\bf x} ~e^{\beta W({\bf x})}$ is a normalisation factor. Eq.~\eqref{eq_thermal_avg} corresponds to an effective thermal average, where the parameters ${\bf x}$ are sampled from a Gibbs distribution $e^{\beta W({\bf x})} / Z_{\beta}$, in which $-W({\bf x})$ plays the role of the energy, and $1/\beta$ is the temperature. We have that \cite{simulated_annealing}:
\begin{equation}
	\langle W \rangle_\beta \underset{\beta \to \infty}{\longrightarrow}  W_{\rm max} ~.
\end{equation}
In words, progressively ramping up the inverse temperature $\beta$, the thermal average Eq.~\eqref{eq_thermal_avg} converges towards the global maximum of the function $W$. At each value of $\beta$, the Gibbs distribution in Eq.~\eqref{eq_thermal_avg} is sampled via a Markov-chain Monte Carlo algorithm \cite{metropolis1953}. 

{\bf Monte Carlo sampling.--} The Markov chain is a list of samples $\{{\bf x}_i\}_{i=1}^N$, generated in such a way that in the limit of infinitely many samples, the average value of $W(\{{\bf x}_i\})$ over the $N$ samples converges towards the exact average value, up to $O(1/\sqrt{N})$ corrections:
\begin{equation}
	\frac{1}{N}\sum_{i=1}^N W({\bf x}_i) = \langle W \rangle_\beta + O(1/\sqrt{N}) ~.
\end{equation}
In order to sample ${\bf x}$ according to the Gibbs distribution [Eq.~\eqref{eq_thermal_avg}], we implement a Metropolis algorithm \cite{metropolis1953}. That is, we start from an arbitrary initial configuration ${\bf x}_1$, and then iterate: \begin{enumerate}
		\item propose a new configuration ${\bf x}_{\rm new}$ (see below)
		\item compute the difference $\Delta = \beta[W({\bf x}_{\rm new}) - W({\bf x}_i)]$
		\item if $\Delta > 0$, accept the move
		\item if $\Delta < 0$, accept the move with probability $e^{\Delta}$
		\item if the move is accepted, update ${\bf x}_{i+1} = {\bf x}_{\rm new}$; otherwise ${\bf x}_{i+1}={\bf x}_i$.
\end{enumerate}

{\bf Implementation of the updates.--}
In our implementation, we ramp $\beta$ linearly from $\beta_i=1$ to $\beta_f \approx 10^4$--$10^5$ in $n_{\rm steps} = 10^3$ steps. For each value of $\beta$, we attempt $n_{\rm attempts}=10^5$ Metropolis updates. Each move consists of selecting randomly one basis among $n$, and rotate randomly all its elements. Specifically, the moves are proposed as follows 
\begin{enumerate}
    \item choose randomly and uniformly one of the bases $y \in [n]$
    \item draw $2d^2$ independent random numbers $\{(r_{jk}, s_{jk})\}_{(j,k) \in [d]^2}$, uniformly in the interval $[-\epsilon, \epsilon]$ (see below for the choice of $\epsilon$)
    \item define $U^y_{\rm new}$ as $U_{jk}^y + r_{jk} + i s_{jk}$
    \item make $U^y_{\rm new}$ unitary via the Gram--Schmidt procedure.
\end{enumerate}
The parameter $\epsilon$, which defines the typical amplitude of the proposed moves, is adapted throughout the algorithm in order to ensure a constant acceptance rate. Intuitively, when the temperature is very high, large moves involving a potentially large change in energy are required to efficiently explore the parameter space. Progressively ramping down the temperature, the bases start to stabilise in the vicinity of the maxima of $W$, and large moves become often rejected by the Metropolis rule. On the other hand, if $\epsilon$ is very small, the moves will not efficiently explore the parameter space. As a compromise, we adapt $\epsilon$ such that the acceptance rate $r_{\rm accept}$ of the Metropolis update (that is, for each value of $\beta$, $r_{\rm accept}$ is the number $N_{\rm accept}$ of accepted moves divided by the number $N_{\rm attempts}$ of attempts), is kept between $r_{\min} = 0.32$ and $r_{\max} = 0.48$. We initialise $\epsilon=1$ at the beginning of the simulation, and whenever $r_{\rm accept}<r_{\min}$, we change $\epsilon$ to $0.8\epsilon$. Similarly, whenever $r_{\rm accept}>r_{\max}$, we change $\epsilon$ to $1.2\epsilon$. 

{\bf Optimal bases.--}
As the goal of the simulation is not to accurately estimate the thermal average $\langle W \rangle_\beta$ [Eq.~\eqref{eq_thermal_avg}], but only to efficiently find its global maximum, we do not carry out a detailed evaluation of the error on $\langle W \rangle_\beta$ as estimated from our samples. Throughout the simulation, we record the optimal bases ${\bf x}_{\rm opt}$ encountered, corresponding to the maximal value of $W_{\rm opt}$ found so far. As a last step of the optimisation, we set $\beta=\infty$ and start a new simulation starting from ${\bf x}_{\rm opt}$ as the initial configuration. Effectively, this amounts to only accepting the moves that increase $W$, in order to achieve as many digits of precision as needed for $W_{\rm opt}$, as well as for the basis elements themselves.

\begin{figure}
    \centering
    \includegraphics[width=\linewidth]{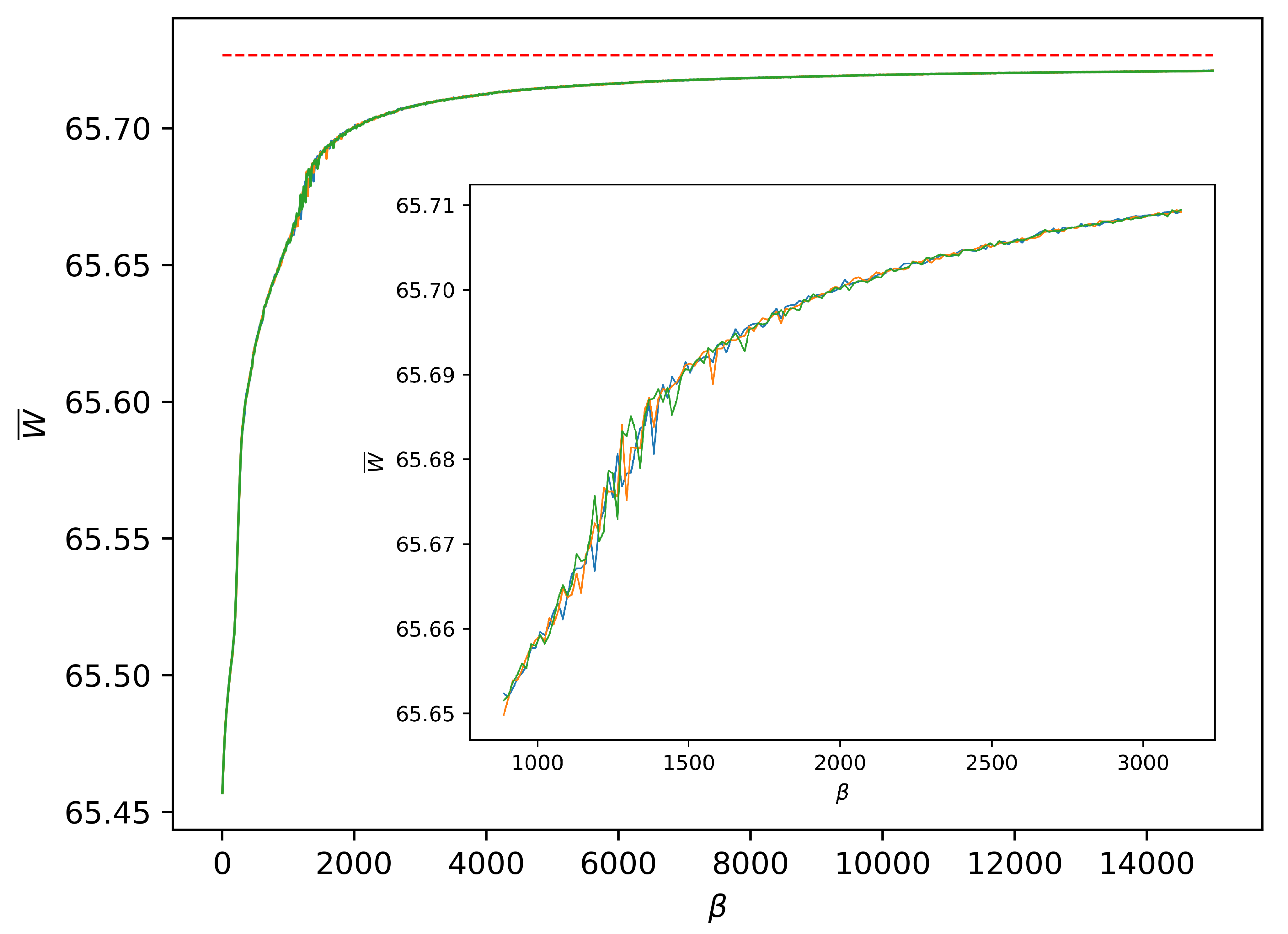}
    \caption{Monte Carlo simulation for $d=6$, $n=4$. As a function of the inverse temperature $\beta$, mean value of the Bell operator [cf.~Eq.~\eqref{eq_thermal_avg}]. Three independent simulations are plotted, and the dashed red line indicates the maximal value found throughout the simulations (see text). Inset: Zoom of the main panel for $1000<\beta < 3000.$}
    \label{fig:d6n4}
\end{figure}

\begin{figure}[hbt]
    \centering
    \includegraphics[width=\linewidth]{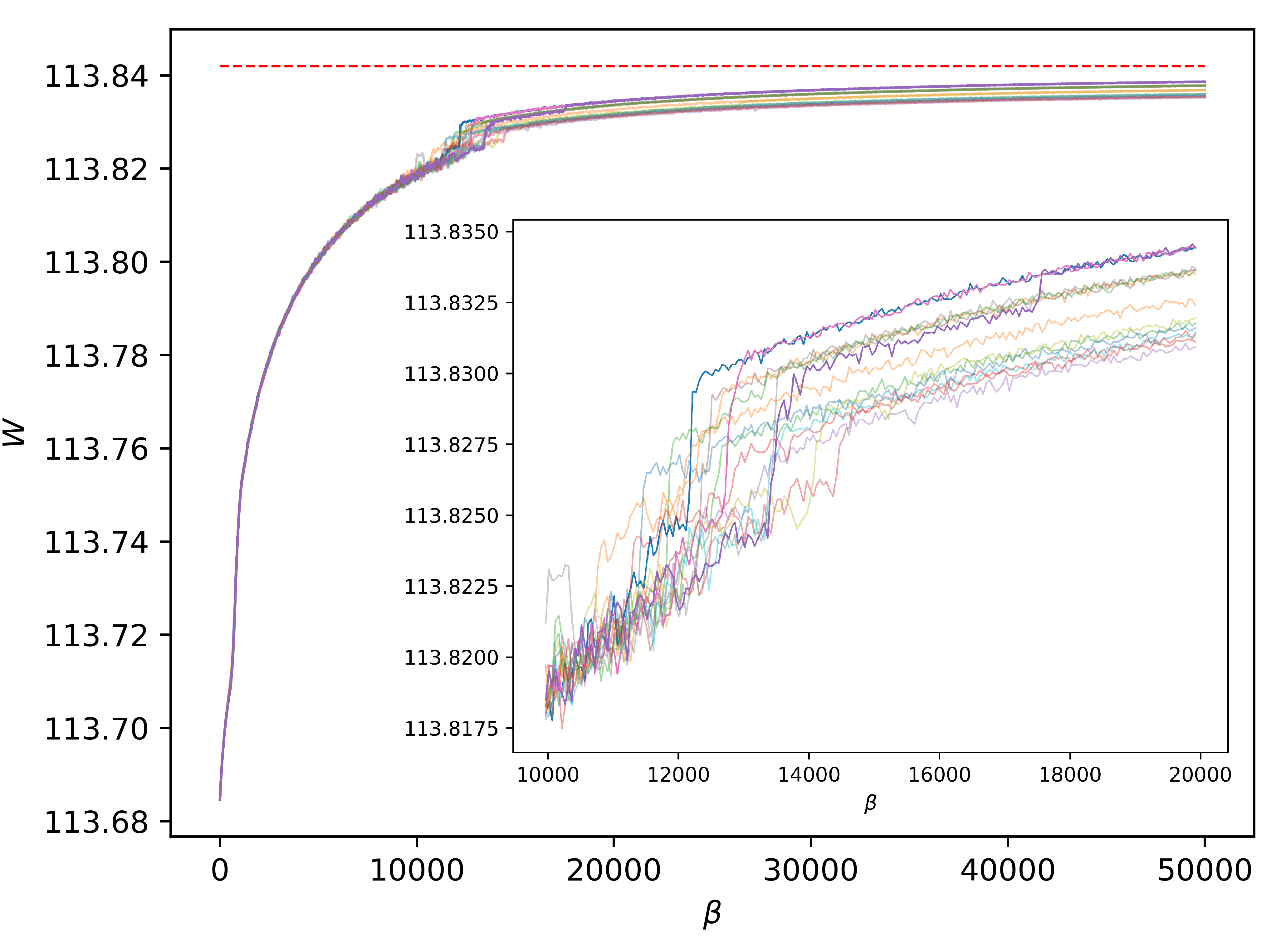}
    \caption{Monte Carlo simulation for $d=10$, $n=4$. As a function of the inverse temperature $\beta$, mean value of the Bell operator [cf.~Eq.~\eqref{eq_thermal_avg}]. Ten independent simulations are plotted, and the dashed red line indicates the maximal value found throughout the simulations. Inset: Zoom of the main panel for $10000<\beta < 20000.$}
    \label{fig:d10n4}
\end{figure}

\subsection{Results}
The results of this optimisation are summarised in Table \ref{tbl:montecarloResults}. For a meaningful comparison of different dimensions $d$, and different number of bases $n$, we indicate the relative deviation to the MUB optimum: $1-W_{\rm max} / W_{\rm MUB}$.

{\bf Power-of-prime dimensions.--} 
All dimensions $2 \le d \le 9$ except $d=6$ are powers of prime, for which the maximal number of MUBs is exactly $d+1$. Our numerical simulations are consistent with this fact, and our variational optimisation systematically reconstructs $n$ MUBs for all $n \le d+1$ \footnote{For $d=8$ we actually stopped at $n=8$, and for $d=9$ at $n=5$, because of the high numerical cost of these higher-dimensional instances.}.
For $d=2,3,4,5$, we also performed the optimisation for $n=d+2$, where three independent simulations gave the same optimum within numerical accuracy, also in agreement with the see-saw and non-linear SDP methods, and with the fact that there cannot exist $d+2$ MUBs in dimension~$d$. The corresponding optimal solutions are analysed analytically in Section \ref{sec:analytic} for $d=2$ and 3.\\

{\bf d=6.--} In the case $d=6$, we do not find more than $n=3$ MUBs. For $n=4$, three independent simulations gave consistently the optimum $W_{\rm opt}(6,4) = W_{\rm MUB}(1 - 0.00004215) \approx 65.723938549\dots$ (within numerical accuracy). The optimal bases found also coincide within numerical precision with those found by the see-saw and non-linear SDP techniques of the previous sections (also see Section \ref{subsec:d6n4} for a close analytical construction). The complete evolution of $\langle W \rangle_\beta$ is illustrated in Fig.~\ref{fig:d6n4} for these three simulations.\\

{\bf d=10.--}
The simulations for $d=10$ are reaching the limits of our current implementation. We did find $n=3$ MUBs, but not $n=4$. As illustrated in Fig.~\ref{fig:d10n4} over ten independent simulations, three of them converged to the same optimum $W_{\rm opt}(10,4) = W_{\rm MUB}(1 - 0.000005938) \approx 113.8413197\dots$. This value is reported in Table~\ref{tbl:montecarloResults}, and is our best estimate for the true optimum\footnote{Using our open source code, Markus Grassl was able to run the simulation for $d=10$ and $n=4$ on a computer cluster 2000 times. He reported to us that the most frequent optimum in these runs agrees with our findings, however, in two instances he found the slightly larger value of $W_{\rm opt}(10,4) \approx 113.8414358$.}. Even though some of our simulations are trapped in local optima, our results support the conjecture that no more than three MUBs exist in dimension 10.

\section{Analytic constructions}
\label{sec:analytic}

In this section we describe analytic constructions that match the best bases found numerically for the cases of four bases in dimension two, five bases in dimension three, and four bases in dimension six.

\subsection{Dimension two}

One can parametrise any rank-1 qubit projection $B$ using the Bloch representation
\begin{equation}\label{eq:Bloch}
B = \frac12( \I + \vec{r} \cdot \vec{\sigma} ),
\end{equation}
where $\vec{r} = (x,y,z)$ is a unit vector in $\mathbb{R}^3$, $\vec{\sigma} = (\sigma_x, \sigma_y, \sigma_z)$ is a vector of the Pauli matrices, and $\vec{r} \cdot \vec{\sigma} = x \sigma_x + y \sigma_y + z \sigma_z$. Accordingly, for the case of four rank-1 projective measurements on a qubit, we use the parametrisation $B^y_1 = \frac12( \I + \vec{r}_y \cdot \vec{\sigma} )$ for $y \in \{1,2,3,4\}$, and $B^y_2 = \I - B^y_1$. Consider then the four measurements parametrised by the four vectors
\begin{equation}\label{eq:tetrahedron}
\begin{split}
\vec{r}_1 & \left. = \frac{1}{ \sqrt{3} } (1, 1, 1) \right. \\
\vec{r}_2 & \left. = \frac{1}{ \sqrt{3} } (1, -1, -1) \right. \\
\vec{r}_3 & \left. = \frac{1}{ \sqrt{3} } (-1, 1, -1) \right. \\
\vec{r}_4 & \left. = \frac{1}{ \sqrt{3} } (-1, -1, 1), \right.
\end{split}
\end{equation}
defining a regular tetrahedron on the Bloch sphere. It is straightforward to verify that by plugging these measurements into Eq.~\eqref{eq:BellnMUB+B} we get
\begin{equation}
W^{+B}_{2,4}(\{B^y_b\}) = 4( \sqrt{3} + \sqrt{6} ) \approx 16.7262,
\end{equation}
which agrees (after normalisation) with the values in Tables \ref{tbl:seesawResults}, \ref{tbl:kktResults}, and \ref{tbl:montecarloResults} up to numerical precision.

\subsection{Dimension three}

For the three-dimensional case, we parametrise each basis $y \in \{1,\ldots,5\}$ by a unitary matrix $U_y$, whose columns correspond to the basis vectors. Fixing the computational basis allows us to write the first basis as $U_1 = \I$. We then take the second basis to be the Fourier basis
\begin{equation}
U_2 = \frac{1}{ \sqrt{3} }
\begin{bmatrix}
1 & 1 & 1 \\
1 & \omega_3 & \omega_3^2 \\
1 & \omega_3^2 & \omega_3
\end{bmatrix},
\end{equation}
where $\omega_3 = \ee^{ \frac{ 2 \pi \ii }{3} }$.  We define the remaining three bases as
\begin{equation}
\begin{split}
U_3 & \left. = \diag( \omega_6, \bar{\omega}_6, 1 ) U_2 \right. \\
U_4 & \left. = \diag( 1, \omega_6, \bar{\omega}_6 ) U_2 \right. \\
U_5 & \left. = \diag( \bar{\omega}_6, 1, \omega_6 ) U_2, \right.
\end{split}
\end{equation}
where $\omega_6 = \ee^{ \frac{ 2 \pi \ii }{6} }$, $\bar{\omega}_6$ is its complex conjugate, and $\diag(x,y,z)$ is a diagonal matrix of $x$, $y$, and $z$. Note that all the bases $U_2$, $U_3$, $U_4$, and $U_5$ are unbiased to $U_1$. It is straightforward to verify that by plugging the measurements $B^y_b$ corresponding to the bases $U_y$ into Eq.~\eqref{eq:BellnMUB+B}, we get
\begin{equation}
W^{+B}_{3,5}(\{B^y_b\}) = 8 ( \sqrt{2} + \sqrt{5} + \sqrt{6} ) \approx 48.7982,
\end{equation}
which agrees (after normalisation) with the values in Tables \ref{tbl:seesawResults}, \ref{tbl:kktResults}, and \ref{tbl:montecarloResults} up to numerical precision.

\subsection{Dimension six}\label{subsec:d6n4}

All three numerical methods converged to (numerically) the same set of four bases in dimension six. Just like in the case of five bases in dimension three, this set has the property that one basis is unbiased to the other three bases. Upon closer inspection, one finds that these bases are numerically very close to the ``four most distant bases'' of Ref.~\cite{RLE11}, which were found via maximising the MUBness measure
\begin{equation}
    D^2 \propto \sum_{(y,z)\in {\rm Pairs}[n]} \sum_{j,k} |\langle b_j^y | b_k^z \rangle|^2 (1 - |\langle b_j^y | b_k^z \rangle|^2) ~.
    \label{eq:MUBness_measure_RLE11}
\end{equation}

One can again parametrise the bases by four unitary matrices such that the first one is $U_1 = \I$. The rest of the bases are based on Eq.~(6) of Ref.~\cite{RLE11}, in which the authors describe a family of three unitary matrices, depending on two parameters, $\theta_t$ and $\theta_x$. The ``optimal'' values of these parameters (optimality here originally means maximising the distance measure of Ref.~\cite{BBEL+07}) is then determined by finding the unique real solution of Eq.~(20) in Ref.~\cite{RLE11}, plugging it into Eq.~(19) of Ref.~\cite{RLE11} to obtain $\theta_t$, and plugging it into Eq.~(21) of Ref.~\cite{RLE11} to obtain~$\theta_x$. These analytic values correspond to approximately $(\theta_x,\theta_t) \approx  (0.9852276,1.0093680)$.

If we substitute the resulting bases $\{ B^y_b \}$ into our optimisation problem in Eq.~\eqref{eq:optimisation_+B}, we obtain (analytically)
\begin{equation}
W_{6,4}^{+B}(\{B^y_b\}) \approx 65.7239381
\end{equation}
Comparing with our optimum $\approx 65.7239385$, we observe that: 1) the analytical optimum (for the bases of Ref.~\cite{RLE11}) and our numerical optimum differ after the eighth significant figure; and 2) consistently, our numerical optimum is larger than the analytical one. This should not be a surprise, for the two solutions optimise different MUBness measures, respectively Eq.~\eqref{eq:MUBness_measure} and Eq.~\eqref{eq:MUBness_measure_RLE11}. As a further comparison, one can look at the overlaps $\tr( B^y_j B^z_k) = |\braket{b^y_j}{b^z_k}|^2$ of the bases found. We indeed find that one of the bases is unbiased to the other three, i.e.
\begin{equation}
\tr( B^1_j B^y_k ) = |\braket{b^y_j}{b^z_k}|^2 = 0.166667 \quad \forall j,y,k.
\end{equation}
Apart from these, there are three different values of overlaps, whose approximate values $0.124$, $0.181$, and $0.152$ agree with our numerical findings up to two or three significant digits.

In conclusion, our three numerical methods, together with the approach of Ref.~\cite{RLE11} all appear to converge to essentially the same set of four bases in dimension six. Small differences in the overlaps originate in the different MUBness measures which are optimised. %might indicate that these ``four most distant'' bases are indeed optimisers of the various ``MUB-ness measures'', providing 
%This provides further numerical evidence supporting Zauner's conjecture.
The results of these four independent different numerical approaches may signify that Zauner's conjecture is indeed correct, although they cannot rule out the contrary.

\section{Conclusions}

We reformulated the existence problem of MUBs as an optimisation problem, using a recently found family of Bell inequalities. We then applied three numerical methods suitable for optimising Bell inequalities in order to tackle the existence problem: see-saw SDP optimisation, non-linear SDP, and Monte Carlo techniques. The results of all these numerical optimisations is in full accordance with the known cases in dimensions $d=2,3,4,5$, where we find $d+1$ MUBs. %whenever $n$ MUBs exist in a given dimension $d$, the algorithms correctly identify these MUBs by finding the appropriate maximum of the corresponding Bell inequality. 
Furthermore, whenever it is known that $n$ MUBs do not exist in a given dimension $d$, %the algorithms converge to a value strictly lower than the hypothetical MUB value. Remarkably, 
all the different algorithms converge to the same set of bases in all cases (with a slight difference between the see-saw method and the other two methods for $d=4$, $n=6$), and these bases are not MUBs.

We applied our numerical techniques to the open case of four MUBs in dimension six. All three algorithms suggest that there do not exist four MUBs in dimension six, by converging to a Bell value strictly smaller than the hypothetical MUB value. Moreover, the bases found by all three algorithms are very close numerically with the ``four most distant bases'' in dimension six of Ref.~\cite{RLE11}. Hence, our findings provide further numerical evidence for Zauner's conjecture. In the next composite dimension, $d=10$, our Monte Carlo results suggest that no more than $n=3$ MUBs exist.

It is important to point out that the numerical methods used in this work are heuristic, i.e.~there is no guarantee of convergence to the global optimum. As such, heuristic numerics can never provide a rigorous proof of the non-existence of MUBs (only that of existence, by explicitly finding MUBs). To overcome this shortcoming, a plausible future direction towards a rigorous numerical proof is using a variant of the Navascu\'es--Pironio--Ac\'in hierarchy of SDPs \cite{NPA07} for maximising a Bell inequality in a fixed dimension. While such numerical optimisation is computationally significantly more expensive than those in our work, it provides certifiable upper bounds on Bell inequality violations, and therefore could in principle be used to rigorously prove the non-existence of MUBs.\\

\noindent\textbf{Code availability.}
The numerical findings presented in this paper can be reproduced using the codes made available on public repositories. For the see-saw SDP, see  \url{https://github.com/Lumorti/seesaw}. For the non-linear SDP, see \url{https://github.com/Lumorti/nonlinear}. For the Monte Carlo simulations, see \url{https://github.com/mariaprat/mubs-montecarlo}.

\section{Acknowledgements}

The authors thank Markus Grassl for providing a large set of data obtained by running our code on a computer cluster. This  project  has  received  funding  from  the  European  Union’s  Horizon  2020  research and innovation programme under the Marie Skłodowska-Curie grant agreement No.~754510, the Government of Spain (FIS2020-TRANQI, Severo Ochoa CEX2019-000910-S), Fundacio Cellex, Fundacio Mir-Puig, Generalitat de Catalunya (CERCA, AGAUR SGR 1381 and QuantumCAT), the ERC AdG CERQUTE and the AXA Chair in Quantum Information Science, the Agence Nationale de la Recherche (Qu-DICE project ANR-PRC-CES47), the John Templeton Foundation (Grant No. 61835), ICFO and SPIE under a Maria Yzuel Fellowship Award.

% \printbibliography
\bibliographystyle{unsrturl}
\bibliography{refs}

\end{document}